\documentclass{article}
\usepackage[T1]{fontenc} 
\usepackage[utf8]{inputenc} 
\usepackage{ismir,amsmath,cite,url}
\usepackage{graphicx}
\usepackage{color}
\usepackage{amsfonts}
\usepackage{footmisc}

\usepackage{multirow}
\usepackage{makecell}
\usepackage{array}
\usepackage{comment}
\newcolumntype{P}[1]{>{\centering\arraybackslash}p{#1}}
\newcolumntype{M}[1]{>{\centering\arraybackslash}m{#1}}
\usepackage{lineno}

\title{Attention-Based Audio embeddings for Query-by-Example}

\multauthor
{Anup Singh$^{1,2}$ \hspace{1cm} Kris Demuynck$^1$ \hspace{1cm} Vipul Arora$^2$} {\\
 $^1$ IDLab, Department of Electronics and Information Systems, imec - Ghent University, Belgium\\
$^2$ Department of Electrical Engineering, Indian Institute of Technology Kanpur, India\\
{\tt\small \{anup.singh, kris.demuynck\}@ugent.be, vipular@iitk.ac.in}
}

\def\authorname{A. Singh, K. Demuynck, and V. Arora}

\begin{document}

\maketitle
\begin{abstract}
An ideal audio retrieval system efficiently and robustly recognizes a short query snippet from an extensive database. However, the performance of well-known audio fingerprinting systems falls short at high signal distortion levels. This paper presents an audio retrieval system that generates noise and reverberation robust audio fingerprints using the contrastive learning framework. Using these fingerprints, the method performs a comprehensive search to identify the query audio and precisely estimate its timestamp in the reference audio. Our framework involves training a CNN to maximize the similarity between pairs of embeddings extracted from clean audio and its corresponding distorted and time-shifted version. We employ a channel-wise spectral-temporal attention mechanism to better discriminate the audio by giving more weight to the salient spectral-temporal patches in the signal. Experimental results indicate that our system is efficient in computation and memory usage while being more accurate, particularly at higher distortion levels, than competing state-of-the-art systems and scalable to a larger database. 
\end{abstract}
\section{Introduction}\label{sec:introduction}

Audio fingerprinting is the principal component of an audio identification task. Finding perceptually similar audio in a massive audio corpus is computationally and memory expensive. Audio fingerprinting is a technique that derives a content-based audio summary and links it with similar audio fragments in the database. It allows for an efficient and quick search against other audio fragments. There are several possibilities for fingerprinting applications on digital devices, such as smartphones and TVs, that are becoming ubiquitous. Music identification on mobile devices, based on query-by-example, is a common use case in which a user hears a song in a public area and wants additional information about it \cite{Shazam, Soundhound}. The second-screen service is another interesting fingerprinting application gaining attention \cite{howson2011second}. It provides meta information of the broadcast content a user is watching/listening to on their devices. In addition, the broadcast sponsors also get benefits from informing viewers and listeners of their products and services. With the ease of music sharing across digital platforms, there is an increasing need to identify copyrighted content, a task which can also be accomplished by audio fingerprinting.

An audio fingerprinting system faces certain challenges for reliable audio identification in a real-world context. First, it should identify the query audio snippets corrupted with distortions such as background noise and reverberation. Secondly,  the system must recognize audio using a few seconds of the audio snippet, which is crucial for fingerprinting systems embedded in digital devices to provide users with an interactive experience. Lastly, the system should generate fingerprints and search in a database in a computationally and memory-efficient manner.

In the past several decades, many audio fingerprinting systems have been developed for audio retrieval tasks. The fingerprinting method proposed by Wang \textit{ et al.}\cite{wang2006shazam} (Shazam) is widely used. It captures a set of salient peaks in the audio spectrogram, assuming they remain unaffected under audio distortions. Further, these peaks are transformed into hashes to enable a fast search. Philips fingerprinting system \cite{haitsma2002highly} is another widely known approach that generates binary fingerprints based on energy changes across spectral-temporal space. However, this approach is computationally intensive. Another approach, named Waveprint, has been introduced in  \cite{baluja2008waveprint}. Waveprint computes the binary fingerprints using top-k wavelets and subsequently processes them using the Min-Hash technique to obtain the compact representations. This system deploys a fast indexing algorithm known as LSH (locality-sensitive hashing) for an efficient audio search. The system proposed in \cite{shibuya2013audio}  utilizes pseudo-sinusoidal components to derive fingerprints robust against noise and reverberation. These systems, however, rely on handcrafted features that make it difficult to accurately identify the query audio when it is distorted with severe background noise and reverberation. Moreover, these systems require long (>5s) audio queries to deliver accurate results.

Deep learning, particularly unsupervised learning, has recently been introduced in the audio fingerprinting domain. \cite{baez2020samaf} proposed SAMAF which utilizes a sequence-to-sequence autoencoder consisting of LSTM layers. Google’s music recognizer system\cite{gfeller2017now} trains a CNN based on the triplet loss function to generate compact audio fingerprints. Another approach proposed in \cite{chang2021neural} trains a similar encoder as \cite{gfeller2017now} using the contrastive learning framework and performs a maximum inner product search within a minibatch during training. 

Typically, the audio is rich in content and contains irrelevant and redundant information that needs to be suppressed or eliminated to generate discriminative audio embeddings.
This fact gives rise to the question of how to enable a CNN to capture the salient information in the signal and suppress the irrelevant ones. In this work, we seek to address this problem using the attention mechanism, inspired by their success in the audio domain \cite{cakir2017convolutional, wang2019environmental}. To this end, we propose to augment the resnet-like architecture with a channel-wise spectral-temporal attention mechanism. The temporal attention mechanism \cite{luong2015effective} has been widely employed in the recurrent neural networks to reweigh the recurrent output at different time indices and combine them to produce a meaningful feature vector at a particular time index. However, the spatial attention mechanism, i.e. attention on feature dimensions, has not been investigated earlier in the context of robust audio representations. Therefore, with the proposed approach, we aim to learn a multidimensional attention mask applied to the CNN features that assign more weight to the salient spatial patches and vice versa. As a result, we expect that the CNN will generate robust audio embeddings. In addition, our work focuses on performing a comprehensive search for our system to be applicable in audio synchronization tasks.

\section{Proposed Method}

Our work aims to generate embeddings for each audio segment of length $L$, extracted with a hop size of $H$ from an audio track.  

Our method builds on simCLR \cite{chen2020simple}, a simple contrastive framework for learning visual representations. It maximizes the similarity between latent representations corresponding to an image under different augmented views using the contrastive loss function. 

We employed this framework in the audio domain by mapping pairs of spectrograms corresponding to clean audio and the distorted one closer to one other in the latent space. In our work, the clean audio was distorted with signal distortions such as background noise and reverberation. Furthermore, a time offset was added to the distorted audio.

\subsection{Audio Preprocessing}



We randomly select a segment $x_{clean}$ of length $L$ from an audio track and resample it to a sampling rate of $F_s$ for training the neural network model. Note that randomly chosen segments may be mostly silent and may thus not contribute to the training of the model. Moreover, such segments introduce errors in the retrieval process. Therefore, the segments with energy levels lower than a threshold $t$ were discarded.

\subsection{Data Augmentation pipeline}

We generate a positive sample $x_{pos}$ corresponding to each $x_{clean}$ by stochastically applying a sequence of augmentations to $x_{clean}$. The following augmentations are considered: 

\begin{itemize}
    \itemsep0em
    \item \textbf{Time offset}: A temporal shift of up to 40\% of $H$ is added to deal with the potential temporal inconsistencies in the real world situation. 
    \item \textbf{Noise mixing:} A randomly selected background noise is added in the range of 0-25dB SNR.
    \item \textbf{Reverberation:} The audio segment is filtered with a randomly selected room impulse response to simulate the room acoustic environments.
    \item \textbf{SpecAugment:} Two randomly chosen time and frequency maskings are applied in the spectrogram. The width of the mask is set to 0.1\% of their respective axis dimension. Moreover, this augmentation serves as a regularizer to prevent overfitting.

\end{itemize}

\begin{table}[h]
    \centering
    \begin{tabular}{l c c}
    \hline
     Layer & Input size & Output size\\
     \hline
     \textbf{Encoder:} \\
     CNN layer & 1$\times$64$\times$96 & 32$\times$64$\times$96 \\
     ResBlock1 & 32$\times$64$\times$96 & 32$\times$64$\times$96  \\
     ResBlock2 & 32$\times$64$\times$96 & 64$\times$32$\times$48  \\ 
     ... \\
     ResBlock6 & 512$\times$4$\times$6 & 1024$\times$2$\times$3 \\
     Flatten & & 6144 \\
     \textbf{Projection Head:} & $d*i$ & $d*o$ \\
     Conv1D + ELU & 128$\times$48 & 128$\times$32 \\
     Conv1D & 128$\times$32 & 128$\times$1 \\
     \hline
    \end{tabular}
    \caption{The proposed model employs a resnet-like architecture as the encoder and a projection head that maps the encoder output to a low-dimensional latent space. The projection head consists of 2 linear layers, each split into $d$ branches with input size $i$ and output size $o$. The model takes a log Mel spectrogram as the input.
}
    \label{tab:arch}
\end{table}

\subsection{Architecture}

\subsubsection{CNN encoder with Spatial-Temporal Attention}
Residual networks\cite{he2016deep} are characterized by skip connections that circumvent some intermediate layers and merge their input and output. The main advantage of such architectures is to prevent the vanishing gradient problem, which hinders training deep network architectures. Therefore, we propose a resnet-like architecture enhanced by the spatial-temporal attention mechanism that enables the CNN to learn discriminative audio representations.

Our proposed architecture contains a front-end that consists of a CNN layer and a resnet block with a kernel stride of 1x1 to retain the full temporal information since we want the discriminative embeddings to also be able to distinguish between fragments that have a similar timbre but different temporal evolution, for example adjacent audio segments.
Furthermore, the back-end consists of subsequent resnet blocks containing 2 convolutional layers. The spatial dimensions of each block are downsampled using a stride of 2x2, while the depth of each block is designed to be double of the preceding layer. Note that every convolution layer uses kernels of size 3x3, and a ReLU activation and batch normalization follows each convolution layer. The overall structure of the architecture is presented in \tabref{tab:arch}.

CNNs have demonstrated their ability to extract complex features from low-level features, such as the log Mel spectrogram. However, the CNN features are translation-invariant, which implies that the spatial regions are treated equally in the feature map, which may not be useful in the audio context. Therefore, using the attention mechanism, we aim to enhance the interesting spatial patches by assigning more weight to time indices and frequency bands containing salient information and vice versa. 

In order to apply the spectral-temporal attention mechanism, a channel-wise mask is computed and applied to the CNN features $X$ of dimension $C \times F \times T$. We use two different attention weights, denoted by $a^{temp} \in \mathbb{R}^{C \times F \times 1}$ for temporal attention and $a^{spect} \in \mathbb{R}^{C \times 1 \times T}$ for spectral attention that are modeled as:

\begin{equation}
    \mathrm{\bm{a^{temp}} = softmax(\bm{X^TW_{temp}})} , 
\end{equation}

\begin{equation}
   \mathrm{\bm{a^{spect}} = softmax(\bm{X^TW_{spect}})} , 
\end{equation}

where $\bm{W_{temp}}$ and $\bm{W_{spect}}$ are learnable weights. Furthermore, the spectral-temporal attention mask $\bm{A} \in \mathbb{R}^{C \times F \times T}$ is computed using the outer product between attention weights:

\begin{equation}
\mathrm{\bm{A} = \bm{a^{spect}} \otimes \bm{a^{temp}}} \times S
\end{equation}

where, $S$ is a scaling factor to rescale the attention mask to enable easy gradient flow during model training. Finally, the CNN feature map is reweighted using the mask $\bm{A}$ as:
\begin{equation}
    \mathrm{\bm{{X'}} = \bm{A} * \bm{X}}
\end{equation}

\begin{figure}[t]
    \centering
    \includegraphics[width=0.47\textwidth]{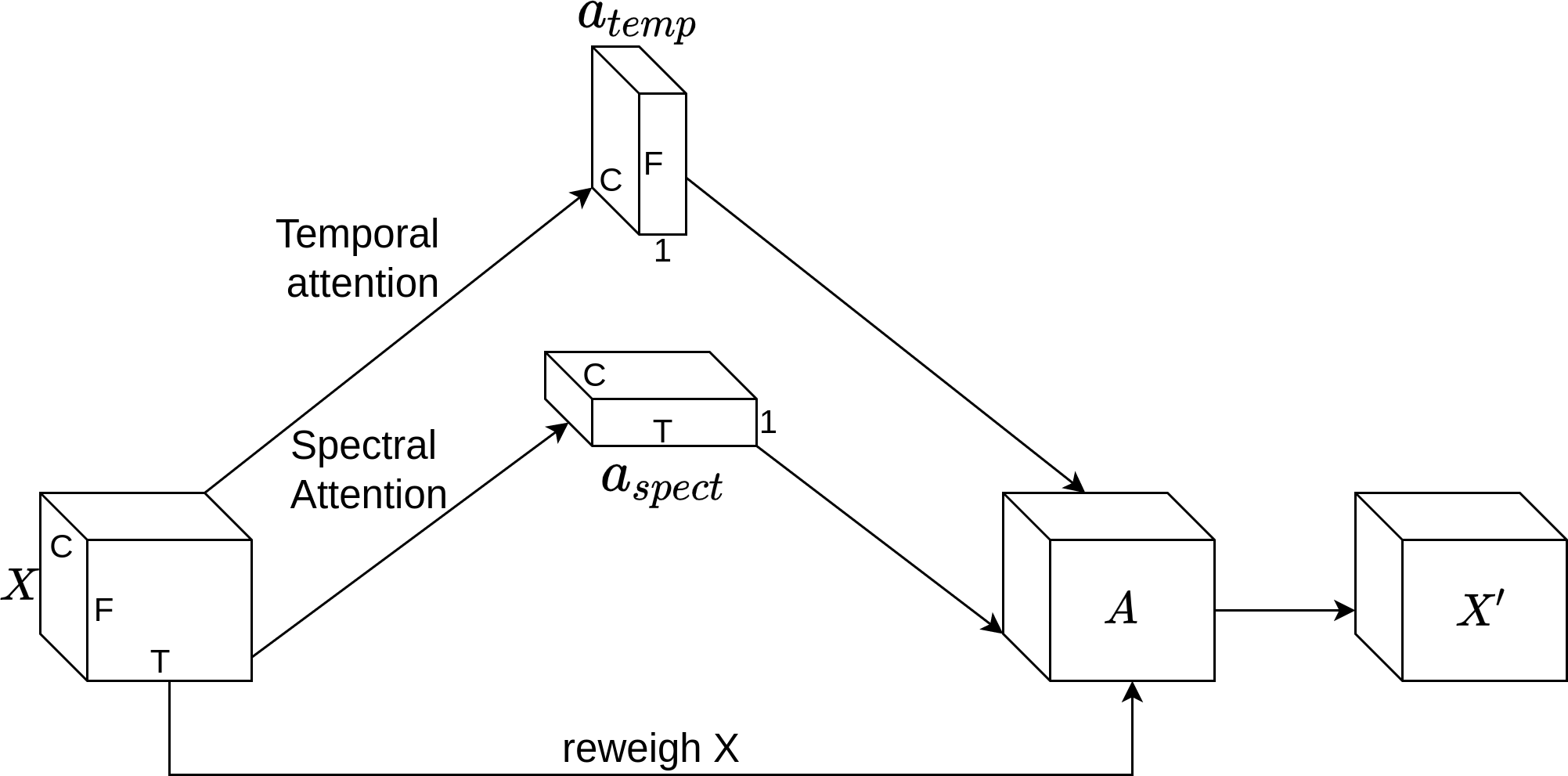}
    \caption{The illustration of spectral-temporal attention mechanism on input feature.}    \label{fig:my_label}
\end{figure}

\subsubsection{Projection Head}
The final output of the CNN encoder is projected into a lower-dimensional latent space via a projection head. The fully connected layers in the projection head result in a large number of model weights. Therefore, as used in previous studies\cite{lai2015simultaneous, gfeller2017now}, we utilize the split head to split the encoder output into $d$ branches. Each branch is then passed through the corresponding linear layers and their final outputs are finally concatenated to generate a $d$-dimensional embedding, followed by  $L^2$ normalization.

\subsection{Contrastive Learning framework}

We employ a contrastive learning approach\cite{chen2020simple} for training the model to map the similar audio samples closer together and pull away different audio samples. As discussed in Section 2.2,  pairs of clean segments and their corresponding distorted versions are generated for training the model. These pairs are equivalent to the anchor and positive sample pairs as denoted in the simCLR framework.  For each sample in a mini-batch of size N (N/2 pairs), there are N-2 negative samples since we also allow a positive sample to be an anchor sample in the batch.
We train the proposed model using these pairs that are mapped into the latent space where the contrastive loss is formulated. We use the normalized temperature-scaled cross-entropy loss as devised in \cite{chen2020simple}: 

\begin{equation}
   \mathrm{ \mathcal{L}_{i,j} = -log \;\frac{\textrm{exp}(sim(e_i,e_j)/\tau)}{\sum_{k=1}^{N-1} 1_{[k \ne i]} \; exp(sim(e_i,e_k)/\tau)}},
\end{equation}

where, $\tau$ is a temperature hyperparameter that allows learning from hard negatives. The cosine similarity is used to calculate pairwise similarity. 


\subsection{Database-creation and Indexing}

The trained model is used as a fingerprinter to extract fingerprints from the audio tracks. The fingerprint for a given audio track is generated as follows: First, we extract segments of length $L$ with a hop size of $H$. These segments are pre-processed as mentioned in Section 2.1. Then, the audio segments are transformed into log Mel spectrograms and fed to the model to generate encoded embeddings. We term these embeddings as subfingerprints, which form a  fingerprint of the entire audio track when stacked in the time–ordering sequence. Finally, we extract the subfingerprints of every audio segment in the database and link them with their corresponding timestamps and the audio identifier to create a reference database.

Finding the nearest neighbors for a sample in a d-dimensional space is not a trivial task (when $\textit{d>50}$)\cite{baluja2008waveprint}. Also, the brute-force search strategy in a massive database is computationally expensive. Therefore, we employ the Locality Sensitive Hashing (LSH)\cite{datar2004locality} to index the subfingerprints database. It allows for efficient retrieval by only comparing a fraction of the dataset to retrieve the exact match. Moreover, it adds the noise-robustness property to the retrieval process.

\subsection{Retrieval process}

This section describes the process of finding the closest matching audio snippet from an indexed database, given a query audio snippet. For a given query audio, we follow the same procedure to generate subfingerprints as mentioned in the previous subsection. Suppose the query $Q=(q_m)_{m=1}^M$ is composed of $M$ subfingerprints. We use LSH to retrieve the nearest match for each subfingerprint $q_m$, denoted as $I_m$ (the index). To identify the closest matching audio track, we retain counts of the audio identifiers corresponding to the retrieved matches. The query audio is finally identified with the audio identifier with the maximum counts. 

We also locate the precise timestamp of the query audio in the identified audio track. To accomplish this, we first eliminate the retrieved matches that do not belong to the identified audio track. Next, we generate the set of candidate sequences $S_i$ of length $M$ with their starting indices as $I_i = I_m - m$. Finally, we select candidate sequence $S_i$, which have at least 50\% intersecting retrieved indices. The timestamp corresponding to $I_i$ is chosen as the matching timestamp of query audio in the retrieved audio track.


\section{Experimental Setup}
\subsection{Dataset}

\begin{itemize} 
    \itemsep0em
    
    \item \textbf{Free Music Archival (FMA)}\cite{defferrard2016fma}: is an open, large-scale, and easily accessible dataset commonly used for various music information retrieval tasks. We used the \textit{fma\_large} and \textit{fma\_medium} versions of the dataset for model training and testing, respectively. There are 106k and 25k 30s audio clips in the \textit{fma\_large} and \textit{fma\_medium}, respectively. Note that we excluded the shared clips between both datasets for training the model. 

    \item \textbf{Noises:} We used a range of real-world background noises for training and testing. We extracted the noise signals from database\footnote{\label{1}\url{https://www.openslr.org/28/}}, which includes a wide range of sounds from the MUSAN corpus\cite{snyder2015musan}, for training the model. For testing the system, we used the ETSI database\footnote{\url{https://docbox.etsi.org/}}, from which we chose seven distinct background noises: Babble, Living Room, Cafeteria, Car, Workplace, Traffic, and Train Station.

    \item \textbf{Room Impulse Responses (RIRs):} We used real RIRs\footref{1} corresponding to different acoustic environments, ranging from a small room to a large hall, for training the model. We selected six RIRs from Aachen Impulse Response Database \cite{jeub09a} with t60 values ranging from 0.1s to 0.8s for testing the system.  
 
\end{itemize}

\subsection{Implementation details}

We compared our approach with a baseline system\cite{chang2021neural} that generates fingerprints using an encoder similar to Now-Playing’s \cite{gfeller2017now} architecture and does a comprehensive search. To the best of our knowledge, it is the only method that employs a neural network model to generate robust audio fingerprints and performs better than conventional approaches. We chose the log Mel spectrograms as input to our model and the baseline method. \tabref{tab:my_label} lists the experiments configurations used for developing our audio fingerprinting system. 

We trained the models with the Adam \cite{kingma2014adam} optimizer for 150 epochs using the cyclic learning rate. The initial learning rate was 5e-4 and reached a maximum of 5e-2 in 40 epochs. We also tweaked the temperature hyperparameter in the [0.01-0.1] range and found no significant benefits. The model was trained on a single NVIDIA Tesla V100 GPU for about 40 hours.  

We built a reference database of $\sim$7.3M fingerprints using the \textit{fma\_medium} dataset. We used LSH implementation\footnote{\url{https://github.com/FALCONN-LIB/FALCONN}} to index the generated audio subfingerprints. Furthermore, the retrieval performance was equivalent to the brute force search after fine-tuning the LSH, with less than a 0.1\% drop in retrieval accuracy at various distortion levels.

\begin{table}[]
    \centering
    \begin{tabular}{l r}
    \hline
        Parameter &  Value \\
    \hline
        Sampling rate ($F_s$)  & 16 kHz \\
        Audio segment length ($L$)  & 960 ms\\
        Energy threshold ($t$) & 0 dB \\
        log Mel spectrogram dimensions (F $\times$ T) & 64 $\times$ 96 \\
        Subfingerprint hop length ($H$)  & 100 ms \\
        Subfingerprint dimensions ($d$)  & 128  \\
        Batch size ($N$)  & 512\\
        Scaling factor ($S$) & 100\\
        \textbf{LSH configuration:} \\
        Tables & 50 \\
        Hash bits & 18 \\
        Number of probes & 200 \\ 
        
     \hline   
        
    \end{tabular}
    \caption{Experiments configurations}
    \label{tab:my_label}
\end{table}

\subsection{Evaluation metric}

We used the following metric for system evaluation at audio/segment level retrieval:

\begin{equation}
    \textit{accuracy} = \frac{\textit{n hits @ top-1}}{\textit{n hits @ top-1} + \textit{n miss @ top-1}} \times \textit{100}
\end{equation}

Note that a match is declared correct for the segment level search if the located timestamp of the query in the correct retrieved audio is within $\pm$ 50 ms.

\subsection{Search query}

We used the \textit{fma\_medium} dataset to generate 10,000 search queries, which were randomly extracted from different audio tracks. To assess system performance at different distortion levels, we distorted queries with noise, reverberation, and a combination of both. To generate a noisy reverberant audio query, we first convolved both the audio and the noise signal with the RIR corresponding to a t60 level of 0.5s and then added both signals at SNR levels ranging from 0dB to 25dB. In addition, we created queries of lengths: 1s, 2s 3s, and 5s to test our system efficiency for varying lengths.   

\section{Results and Discussion}
In the following, we present the performance of our system under different distortion conditions. We compare our system with the baseline system as mentioned in the previous section and the Audfprint system.

\subsection{VS. Baseline system}

\begin{itemize}
    \item \textbf{Noise:} \tabref{tab:noise_res} presents the performance of the systems in noisy conditions. It can be seen that our system performs better than the baseline system by a reasonable margin, particularly at the 0dB and 5dB SNR levels. Moreover, our system can precisely locate the timestamp with reasonable accuracy, given enough query length (>2s) in very high noise conditions too. The performance gap narrows with the increase in SNR level and becomes smaller from 20dB onwards. We noted that the performance gap between systems was less than 2\% at SNR levels of 20dB and more, irrespective of the query lengths.

    \item \textbf{Reverb:} As can be seen in \tabref{tab:rev_r}, the retrieval accuracy of the systems drops with an increase in t60 levels, i.e., high reverberant environments. Furthermore, we discovered that the reverberation causes embeddings correspondings to adjacent audio segments to be very similar, which is most likely the reason that the correct match was not found at the top-1 rank in many cases, which resulted in low retrieval performance of the system. Nevertheless, the presented results indicate that our system performs effectively even in high reverberation environments for short query snippets. The baseline system is quite effective in a low reverberation environment with more than 80\% retrieval accuracy. However, its performance falls short at higher reverberations, even with long query snippets.

    \item \textbf{Noise and Reverb:} The systems were tested in a more challenging situation by considering a noisy and reverberant environment.  As shown in \tabref{tab:nrev_r}, the performance of both systems degrades with the addition of reverberation compared to their performance under noisy conditions. Due to added reverberation, the retrieval accuracy of our system drops by around 15\% at 0dB SNR, but it improves on the less noisy conditions.  On the contrary, the baseline system performs poorly, particularly at 0dB SNR. Furthermore, the accuracy gap remains over 15\%  compared to its performance in noisy conditions, even at higher SNR levels. It indicates that our system is more resilient against noisy and reverberant environments than the baseline system. 
 
\end{itemize}

\begin{table}[h]
    \centering
    \small
    \renewcommand{\arraystretch}{1.5}
    \begin{tabular}{c|M{1.4cm}|M{0.8cm}M{0.8cm}M{0.8cm}M{0.8cm}}
    \hline

     Method   &\centering Query length (s)    & 0dB    & 5dB    & 10dB   & 15dB   \\
     \hline
     \hline
     Ours & \multirow{2}{*}{\centering0.96}& \textbf{76.8} & \textbf{84.8} & \textbf{87.4} & \textbf{88.5} \\
     Baseline & & 62.7 & 79.4 & 85.5 & 87.6\\
     \hline
     Ours & \multirow{2}{*}{2}& \textbf{82.7} & \textbf{90.8} & \textbf{93.2} & \textbf{94.1}  \\
     Baseline & & 71.1& 86.5 & 90.4 & 92.0 \\
     \hline
     Ours & \multirow{2}{*}{3}& \textbf{83.9} & \textbf{92.9} & \textbf{94.6} & \textbf{95.5}  \\
     Baseline & & 76.8 & 88.5 & 91.4 & 93.8 \\
     \hline
     Ours & \multirow{2}{*}{5}& \textbf{85.8} & \textbf{93.9} & \textbf{94.7} & \textbf{96.8}  \\
     Baseline & & 79.8 & 89.4 & 91.5 & 94.2\\
     \hline
     
    \end{tabular}
    \caption{Top-1 hit rate (\%) performance in the segment-level search for varying query lengths in noisy conditions.}
    \label{tab:noise_res}
\end{table}

\begin{table}[h]
    \centering
    \small
    \renewcommand{\arraystretch}{1.5}
    \begin{tabular}{c|M{1.4cm}|M{0.5cm}M{0.5cm}M{0.5cm}M{0.5cm}M{0.5cm}}
    \hline

     Method   &\centering Query length (s) & 0.2s&0.4s& 0.5s &0.7s &0.8s     \\
     \hline
     \hline
     Ours & \multirow{2}{*}{\centering0.96}& \textbf{85.1} & \textbf{84.1} & \textbf{78.8} & \textbf{83.3} & \textbf{74.4} \\
     Baseline & & 78.4 & 75.5 & 67.5 &75.1 &62.2 \\
     \hline
     Ours & \multirow{2}{*}{2}& \textbf{89.1} & \textbf{87.3} & \textbf{80.6} & \textbf{85.4} & \textbf{75.0} \\
     Baseline & &  85.5 & 81.3 & 72.5 & 78.6 & 64.9\\
     \hline
     Ours & \multirow{2}{*}{3}& \textbf{90.1} & \textbf{87.9} & \textbf{81.0} & \textbf{86.8} & \textbf{76.3} \\
     Baseline & & 87.2 & 83.3 & 73.9 & 80.8 & 66.9 \\
     \hline
     Ours & \multirow{2}{*}{5}& \textbf{91.2} & \textbf{89.6} & \textbf{82.6} & \textbf{88.4} & \textbf{77.4} \\
     Baseline & & 87.8 & 84.3 & 75.1 & 81.9 & 67.6\\
     \hline
    \end{tabular}
    \caption{Top-1 hit rate (\%) performance in the segment-level search for varying query lengths in reverberant conditions.}
    \label{tab:rev_r}
\end{table}

\begin{table}[h]
    \centering
    \small
    \renewcommand{\arraystretch}{1.5}
    \begin{tabular}{c|M{1.4cm}|M{0.8cm}M{0.8cm}M{0.8cm}M{0.8cm}}
    \hline
    
     Method   &\centering Query length(s)    & 0dB    & 5dB    & 10dB    & 15dB  \\
     \hline
     \hline
     Ours & \multirow{2}{*}{\centering0.96}& \textbf{60.3} & \textbf{76.6} & \textbf{81.3} & \textbf{82.8} \\
     Baseline & & 27.3 & 58.7& 70.7 & 73.9  \\
     \hline
     Ours & \multirow{2}{*}{\centering2}& \textbf{66.4} & \textbf{83.5} & \textbf{86.9} & \textbf{88.0} \\
     Baseline & & 39.0 & 69.6& 76.5 & 78.7  \\
     \hline
     Ours & \multirow{2}{*}{3}& \textbf{67.9} & \textbf{85.1} & \textbf{88.2} & \textbf{89.3}\\
     Baseline & & 47.1 & 75.2 & 80.2 & 81.4 \\
     \hline
     Ours & \multirow{2}{*}{5}& \textbf{69.5} & \textbf{87.1} & \textbf{90.5} & \textbf{91.9}  \\
     Baseline & & 54.7& 77.3 & 81.8 & 82.8 \\
     \hline
    \end{tabular}
    \caption{Top-1 hit rate (\%) performance in the segment-level search for varying query lengths in noisy reverberant conditions.}
    \label{tab:nrev_r}
\end{table}

The above-stated results show that our system performs reasonably well with short query snippets in different distortion environments. Furthermore, it indicates that our system does not require long queries to achieve reasonable performance at higher distortion levels. However, the performance of the systems improves with the increase in the query length using our proposed simple yet effective sequence search strategy. Therefore, the system can be fed with longer queries to obtain more reliable results. 

\textbf{Attention mechanism effect:} Based on system performance in a noisy reverberant environment, we examine the effectiveness of adding an attention mechanism to the model. \tabref{tab:attention} shows that the attention mechanism improves retrieval accuracy, particularly at low SNR levels, with small benefits at higher SNR levels. Furthermore, we noted similar results hold true in other distortion conditions, i.e., noisy and reverberant environments. These results support that the spatial-temporal attention mechanism enhances the CNN to generate robust audio embeddings.

\begin{table}[]
    \centering
    \begin{tabular}{c|ccccc}
    \hline
    & 0dB & 5dB & 10dB & 15dB & 20dB \\
     \hline
     \hline
         Attention &  60.3 & 76.6 & 81.3 & 82.8 & 84.6 \\
         No attention & 52.4 & 68.1 & 75.7 & 79.9 & 82.3 \\
         \hline
    \end{tabular}
    \caption{The effect of added attention mechanism on Top1-hit rate performance of the system in noisy reverberant environments for 0.96s long queries. }
    \label{tab:attention}
\end{table}

\textbf{Embedding dimensions:} We examined the effect of the audio embedding dimension on the system performance using the 0.96s queries.  We observe that shrinking the dimensions from 128 to 64 has little impact on the system, especially in the reverberant environment. However, in the noisy reverberant and noisy environments, retrieval accuracy declines by 5.6\% and 3.8\% at 0dB SNR, respectively, whereas accuracy losses are minimal at other SNR levels. Furthermore, increasing the number of dimensions from 128 to 256 does not provide significant improvements. This investigation allows reducing dimensions further without a significant performance drop to resolve the space constraints problem.

\textbf{Computational and Memory load: }
We further investigated the efficiency of our system based on its computational and memory requirements. The final size of the subfingerprints database is roughly 1.25GB, with 128 32-bit
floating numbers representing each audio subfingerprint. We use the \textit{Intel Xeon Platinum 8268} CPU to do an in-memory search due to the small database size. We report that the LSH takes about 0.01s to retrieve top-5 matches for a query subfingerprint. Moreover,  our system takes about 0.25s to process a 3s long query and locate its timestamp in the identified reference audio. It is also worth noting that the retrieval process can be sped up by employing a parallel search method. Furthermore, there is scope for investigating the computation load reduction by storing subfingerprints in low-bit floating numbers without a significant drop in the retrieval accuracy.


\subsection{VS. Audfprint}
We also compare our system with the Audfprint\footnote{\label{4}\url{https://github.com/dpwe/audfprint}} based on the Shazam method. We notice that Audfprint performs poorly in the segment level search, particularly at high distortion levels. Therefore, we only compare its performance with ours for the audio identification task. Moreover, its performance deteriorates with short query lengths; hence we present the retrieval results with 5s audio query snippets in \tabref{tab:shazam}. 

It can be seen that our system delivers excellent retrieval accuracy, with over 95\% accuracy under high distortion conditions. On the contrary, the Audfprint method performance degrades severely at high distortion levels, which indicates that our system also outperforms the conventional approach for the audio identification task.  It should be noted that the Audfprint system uses a hash table to index the database, resulting in reduced fingerprint database size. Furthermore, the size of the database generated by Audfprint is around 400 MB, which is roughly three times less than ours.

\begin{table}[h]
    \centering
    \small
    \renewcommand{\arraystretch}{1.5}
    \begin{tabular}{c|M{1.25cm}|M{0.8cm}M{0.8cm}M{0.8cm}M{0.8cm}}
    \hline
    Distortion & Method & 0dB & 5dB & 10dB &15dB\\
    \hline
    \multirow{2}{*}{Noise} & Ours &  \textbf{95.0} & \textbf{98.7} & \textbf{98.9} & \textbf{99.2} \\
    & Audfprint & 72.1 & 82.7 & 89.4 & 91.2\\
    \hline
    \multirow{2}{*}{\makecell{Noise+\\Reverb}} & Ours &  \textbf{84.3} & \textbf{96.8} & \textbf{98.5} & \textbf{98.9} \\
    & Audfprint & 64.8 & 79.4 & 87.2 & 92.3 \\
    \hline
    \end{tabular}
    \centering
    \small
    \renewcommand{\arraystretch}{1.5}
    \begin{tabular}{c|p{1.24cm}|M{0.55cm}M{0.56cm}M{0.56cm}M{0.55cm}M{0.55cm}}
    \hline
    \hspace{1.3cm} & \hspace{1cm} & 0.2s & 0.4s & 0.5s & 0.7s & 0.8s\\
     \hline
     \multirow{2}{*}{Reverb} & \centering Ours &  \textbf{99.2} & \textbf{99.5} & \textbf{98.9} & \textbf{99.6} & \textbf{98.7} \\
    & Audfprint & 96.1 & 94.6 & 81.8 & 89.6 & 40.2\\
    \hline
    \end{tabular}
    \caption{Top-1 hit rate (\%) performance in the audio-level search in different distortion conditions.}
    \label{tab:shazam}
\end{table}

\section{Conclusion}
This paper presents an audio fingerprinting system robust against high noise and reverberation conditions. Our work focuses on generating robust audio embeddings by employing a contrastive learning framework. Moreover, we propose to enhance CNN with the channel-wise spectral-temporal attention mechanism to reweigh the CNN features. This enables CNN to assign more weight to salient patches in the CNN features, resulting in discriminative audio embeddings. Furthermore, our system performs a comprehensive search to precisely estimate the timestamp of the query in the identified reference audio using a simple sequence search strategy, which makes our system applicable to audio synchronization tasks. Our system performs well compared to the baseline \cite{chang2021neural} and Audfprint\footref{4} methods. Also, our system is computationally and memory-efficient due to the compact embeddings that make our system deployable on an extensive database. The future direction of this work is to obtain discrete audio embeddings to speed up the audio retrieval process.

\section{Acknowlegment}

This work has been supported by the research grant(PB/EE/2021128-B) from Prasar Bharati. We would also like to extend our gratitude to IITK Paramsanganak for their GPU computing resources.

\bibliography{ISMIRtemplate}

%
%
%
%
%

\end{document}